\documentclass{bmcart}

\usepackage[utf8]{inputenc} 

\usepackage{amsmath, amsfonts}
\usepackage{graphicx}
\usepackage{bbm} 
\newcommand{\tss}{\textsuperscript} 

\usepackage{array} 
\newcolumntype{C}[1]{>{\centering\let\newline\\\arraybackslash\hspace{0pt}}p{#1}}
\usepackage{color}

\begin{document}

\begin{frontmatter}

\begin{fmbox}
\dochead{Research}


\title{Semi-supervised graph labelling reveals increasing partisanship in the United States Congress}


\author[
   addressref={UofA,ACEMS},                   
   corref={UofA},                       
   email={max.glonek@adelaide.edu.au}   
]{\inits{M}\fnm{Max} \snm{Glonek}}
\author[
   addressref={UofA,ACEMS},
]{\inits{J}\fnm{Jonathan} \snm{Tuke}}
\author[
   addressref={UofA,ACEMS,D2D},
]{\inits{L}\fnm{Lewis} \snm{Mitchell}}
\author[
   addressref={UofA,ACEMS},
]{\inits{N}\fnm{Nigel} \snm{Bean}}


\address[id=UofA]{
  \orgname{School of Mathematical Sciences, University of Adelaide}, 
  \city{Adelaide},                              
  \state{SA},
  \postcode{5005},                                
  \cny{Australia}                                    
}
\address[id=ACEMS]{%
  \orgname{ARC Centre of Excellence for Mathematical and Statistical Frontiers}
}
\address[id=D2D]{
\orgname{Stream Lead: Beat the News, Data to Decisions CRC}
}


\end{fmbox}


\begin{abstractbox}

\begin{abstract}
Graph labelling is a key activity of network science, with broad practical applications, and close relations to other network science tasks, such as community detection and clustering. While a large body of work exists on both unsupervised and supervised labelling algorithms, the class of random walk-based supervised algorithms requires further exploration, particularly given their relevance to social and political networks. This work refines and expands upon a new semi-supervised graph labelling method, the GLaSS method, that exactly calculates absorption probabilities for random walks on connected graphs. The method models graphs exactly as discrete-time Markov chains, treating labelled nodes as absorbing states. The method is applied to roll call voting data for 42 meetings of the United States House of Representatives and Senate, from 1935 to 2019. Analysis of the 84 resultant political networks demonstrates strong and consistent performance of GLaSS when estimating labels for unlabelled nodes in graphs, and reveals a significant trend of increasing partisanship within the United States Congress.
\end{abstract}




\end{abstractbox}
%

\end{frontmatter}



\section*{Introduction}
\label{s1}
Graph labelling is concerned with the problem of estimating the labels of one or more nodes within a graph, where an association between the graph's structure and the distribution of labels is assumed to exist. Many graph labelling algorithms exist, both supervised~\cite{azran-07,hassan-10,talukdar-08} and unsupervised~\cite{perozzi-14,pons-05,zhou-04}. In both approaches, a graph comprises unlabelled and labelled nodes, and the algorithms seek to estimate the labels of the unlabelled nodes. While a diverse range of graph labelling methods exist~\cite{fortunato-10,hamilton-17}, this work focusses on the class of dynamical and statistical inference methods that use random walks.

One prominent application of network science is in the analysis of political networks~\cite{victor-17,ward-11}, including the labelling of nodes in political voting networks. Previous works have examined methods to locate individual politicians within a multidimensional political spectrum~\cite{poole-85,poole-01}, the detection of voting blocs or communities within a political voting network~\cite{waugh-09}, and an analysis of partisanship trends reflected in voting networks from the United States Congress~\cite{andris-15,poole-84}. This work presents an analysis of a large collection of United States Congressional roll call voting networks, using a semi-supervised graph labelling method to determine the party affiliation of individuals. Changes in partisanship over time are also examined, with results in accordance with previous studies~\cite{andris-15,poole-84}.

\subsection*{Related Work}
\label{ss1.1}
\subsubsection*{Random Walk-Based Graph Labelling Methods}
\label{sss1.1.1}
In unsupervised algorithms, the graph is organised into clusters, without consideration of the labelled nodes. Once clustered, labels for unlabelled nodes in the graph can be estimated based on the clusters to which labelled nodes belong. However, cases may arise where an identified cluster contains no labelled nodes, or where a cluster contains multiple nodes with different labels, creating uncertainty as to how labels should be estimated for nodes in such clusters.

The Walktrap algorithm is one commonly used random walk-based unsupervised graph labelling method ~\cite{pons-05}. Walktrap searches for densely connected subgraphs by simulating short random walks on a graph, reasoning that short walks are more likely to remain in the same cluster than to leave it. Walktrap quantifies the similarity between nodes using a distance metric, then recursively merges identified clusters based on short random walks, providing a hard classification for each node. Because Walktrap does not use information about labelled nodes, there is no generally accepted method for estimating the labels for unlabelled nodes based on the clusters it identifies.

A more recent unsupervised graph labelling method, DeepWalk, combines random walks and natural language processing tools~\cite{perozzi-14}. DeepWalk uses short, truncated random walks to explore the neighbourhood for every node in a graph, as a means of community detection. The results of these random walks are then combined with a neural language model to generate a low-dimensional representation of the graph. The latent community membership information encoded by this low-dimensional embedding can then be used for multi-label classification tasks.

Unlike unsupervised algorithms, supervised algorithms utilise the information contained in labelled nodes when estimating the labels of unlabelled nodes. A common approach is to treat labelled nodes as absorbing states and unlabelled nodes as transient states in a discrete-time Markov chain (DTMC), and estimate the absorption probabilities or expected times to absorption for all transient states in the chain. Labels for each unlabelled state can then be estimated using the approximate probabilities or times. However, while existing supervised and semi-supervised methods use both labelled nodes and the graph's structure to estimate labels, they only approximate absorption probabilities and times, rather than calculating them exactly.

The Rendezvous algorithm~\cite{azran-07} labels nodes in a semi-supervised setting by constructing a simplified, ``rendezvous'' graph, where edges are drawn from an unlabelled node to only its \(M\) nearest neighbours. \(M\) is chosen to be as small as possible while ensuring that each unlabelled node in the rendezvous graph is connected to at least one labelled node. Once the rendezvous graph has been constructed, edge weights are calculated using a Euclidean distance metric, and absorption probabilities are calculated using the eigenvalues and eigenvectors of the rendezvous graph's transition matrix. Absorption probabilities for nodes in the rendezvous graph are then used to estimate the labels of nodes in the full graph.

Another semi-supervised graph labelling method seeks to label nodes in a binary setting according to expected time to absorption, rather than absorption probability~\cite{hassan-10}. The ``Censored Time'' method simulates step-limited random walks over a graph, recording the number of steps taken for all walks that are absorbed before being terminated by the step limit. The censored times to absorption for absorbed walks are used to approximate the conditional expected time to absorption in each labelled node in the graph. A ``hard'' binary classification is used to estimate labels according to the lowest censored conditional time to absorption.

\subsubsection*{Political Science and Networks}
\label{sss1.1.2}
Analysis of United States Congressional voting data is a popular activity within the field of political science and political networks, in part because large amounts of voting data are freely available~\cite{vv-18}. Various attempts have been made to analyse voting trends within Congressional voting data, including modelling Congresses as political networks, where nodes represent individual politicians, and edges capture some relationship between them.

DW-NOMINATE~\cite{poole-01}, and its predecessors, D-NOMINATE~\cite{poole-85} and \linebreak W-NOMINATE, represents one of the most detailled attempts to study voting behaviour and trends in the United States Congress. A multidimensional scaling method, DW-NOMINATE models individual politicians as points embedded in multidimensional space. Each point, representing the politician's true political alignment, can be estimated by analysing historical voting records. Individuals with similar ideologies (as reflected by their voting records) are spatially ``close'' to one another, while individuals with differing ideologies are ``distant''. Amongst many other applications, DW-NOMINATE is notable for its use in analysing changes in partisanship over time~\cite{poole-84}.

More recent work also discusses changes in partisanship in roll call voting networks over time~\cite{andris-15}. The work examines pairs of nodes within roll call voting networks, modelling the probability distibutions for cooperation between politicians (edges between nodes) from the same party and from opposing parties. A significant long-term trend of increasing partisanship and decreasing inter-party cooperation is identified; increasing the probability of edges between nodes from the same party and decreasing the probability of edges between nodes from opposing parties. The work also makes reference to the continued, though diminishing, presence of ``super-cooperators'' - members who cooperate across party lines - in Congress.

Separate work examining United States Congressional voting data uses modularity to measure political polarisation~\cite{waugh-09}. This work detects voting blocs or communities within Congressional roll call voting networks without making assumptions that rely on the two-party system, revealing more (and more varied) groups than simply Democrats and Republicans. The composition and behaviour of blocs is observed to vary significantly over time, as are the strengths of connections between blocs. The work reveals not only increases in partisanship over time, but points to a possible underestimation of partisanship by other methods in Congresses with weaker party structure.

\subsection*{Contributions}
\label{ss1.2}
This work expands upon a new semi-supervised graph labelling method, the Graph Labelling Semi-Supervised (GLaSS) method, using random walks to absorption~\cite{glonek-18}. The method models a graph as a DTMC, where transient states correspond to unlabelled nodes, and absorbing states correspond to labelled nodes. The transition matrix \(P\), for the DTMC, is formed from the graph's weighted adjacency matrix by normalising the weighted out-degree of each node in the network. From careful construction of \(P\), the probability of absorption in each absorbing state can be calculated exactly, and these probabilities can then be used to estimate the label for every node corresponding to a transient state in the DTMC.

By calculating exact absorption probabilities and expected times to absorption, the GLaSS method provides better label estimates than contemporary supervised methods, which rely on approximations of these quantites~\cite{glonek-18}. By utilising the information contained in labelled nodes in the graph, GLaSS also provides a clear method for estimating the label of unlabelled nodes using quantities that are meaningful and interpretable, unlike unsupervised random walk methods.

This work also contributes to existing work on political networks, through the analysis of a large collection of US Congressional roll call voting networks. In particular, this work contains the first analysis of roll call voting networks using a random walk-based graph labelling method, while also identifying notable trends within the House of Representatives and the Senate. The GLaSS method is able to detect and confirm rising partisanship within the House of Representatives and the Senate~\cite{andris-15}, while also identifying possible historical periods of reduced partisanship.

This work formally introduces the GLaSS method, describes, in detail, the data to be analysed, presents a full description of all analyses performed, and discusses the results of this analysis and possible areas of further work.

\section*{Method}
\label{s2}
Consider an undirected graph \(G~=~(V,E)\) comprising \(n\) nodes, \(V~=~\{v_{1},...,v_{n}\}\), connected by a set of positive real-weighted edges \(E
\). Define the weighted adjacency matrix \(A = [a_{i,j}]\), where \(a_{i,j}~=~a_{j,i}\) records the weight of the edge connecting \(v_{i}\) and \(v_{j}\), and \(a_{i,j}~=~0\) if no edge connects \(v_{i}\) and \(v_{j}\). Suppose the first \(u\) nodes in \(G\) are unlabelled, and the remaining \(\ell\) nodes in \(G\) are labelled, where \(n~=~u+\ell\), and construct the sets \(U~=~\{1,...,u\}\) and \(L~=~\{u+1,...,n\}\) to index the unlabelled and labelled nodes of \(G\), respectively. Arrange \(A\) as
\[A = \begin{bmatrix}
\; A_{U,U} \; & \; A_{U,L} \; \\
\; A_{L,U} \; & \; A_{L,L} \;
\end{bmatrix},\]
where \(A_{J,K}\) describes the weighted edges connecting nodes indexed by \(J\) to nodes indexed by \(K\).

Consider a random walk on \(G\), described by a discrete-time Markov chain (DTMC) where all unlabelled nodes map to transient states and all labelled nodes map to absorbing states. Let \(X_{t}\) denote the state of the chain at time \(t\). For all transient states in the DTMC (unlabelled nodes in \(G\)), calculate the transition probabilities for the DTMC using the adjacency matrix \(A\), where
\begin{equation}
\label{eqn1}
p_{i,j}= P(X_{t+1} = j \ | \ X_{t} = i) = \frac{a_{i,j}}{\sum_{k=1}^{n}a_{i,k}},\quad\forall~i \in U,~j \in V
\end{equation}
is the probability that the DTMC moves to state \(j\) at the next time step, given that the DTMC is currently in state \(i\). For all absorbing states in the DTMC (labelled nodes in \(G\)), by definition,
\begin{equation}
\label{eqn2}
p_{i,j} =
\begin{cases}
\mbox{\(1\),} & \mbox{if \(j = i\)} \\
\mbox{\(0\),} & \mbox{otherwise}
\end{cases}
,\quad\forall~i \in L,~j \in V.
\end{equation}
Gather these probabilities into sub-matrices \(P_{U,U}\:(i,j \in U)\), \(P_{U,L}\:(i \in U,\:j \in L)\), \(P_{L,U}\:(i \in L,\:j \in U)\), and \(P_{L,L}\:(i,j \in L)\). Note that, by definition, \(P_{L,U} = 0\) and \(P_{L,L} = I\). Then, construct the transition matrix
\begin{equation}
\label{eqn3}
P = [p_{i,j}] = \begin{bmatrix}
\; P_{U,U} \; & \; P_{U,L} \; \\
\; P_{L,U} \; & \; P_{L,L} \;
\end{bmatrix}.
\end{equation}

The \(u \times u\) matrix \(P_{U,U}\) governs transitions between transient states, and the \(u \times \ell\) matrix \(P_{U,L}\) governs transitions from transient states to absorbing states.

\subsection*{DTMC Absorption Probabilities}
\label{ss2.1}
Let \(h_{i,j}\) be the probability that the DTMC is eventually absorbed in state \(j, \forall\:j \in L\), given that the chain starts in state \(i, \forall\:i \in U\). Define the matrix of absorption probabilities \(H~=~[h_{i,j}]\). \(H\) is restricted to have \(u\) rows and \(\ell\) columns, corresponding to the \(u\) transient states and \(\ell\) absorbing states of the DTMC, respectively. Then \(H\) can be formally calculated as the minimal non-negative solution to 
\begin{equation}
\label{eqn4}
(I_{u} - P_{U,U})H = P_{U,L}.
\end{equation}
where \(I_{u}\) is the \(u \times u\) identity matrix~\cite{grinstead-12}.

\subsection*{Semi-Supervised Graph Labelling}
\label{ss2.2}
Given a graph \(G\) and the matrix of absorption probabilities \(H\), let the random variable \(Y_{i}\) be the label of an unlabelled node \(v_{i}\), and let \(x_{j}\) be the label of a labelled node \(v_{j}\). The distribution over \(Y_{i}\) can be directly derived from \(H\), for all \(i~\in~U\), as follows:
\begin{equation}
\label{eqn5}
P(Y_{i} = k) = \sum_{j \in L}h_{i,j}\mathbbm{1}(x_{j} = k)
\end{equation}
where \(\mathbbm{1}\) is the indicator function, taking value \(1\) if its argument is true, and \(0\) otherwise.

\subsection*{DTMC Expected Times to Absorption}
\label{ss2.3}
Let \(t_{i}\) be the expected number of time steps before the DTMC is absorbed in any absorbing state, given that the chain starts in state \(i\). Define the vector of expected times to absorption \(\mathbf{t}~=~(t_{1},...,t_{u})^{T}\), where the \(u\) elements of \(\mathbf{t}\) correspond to the \(u\) transient states of the DTMC. Then \(\mathbf{t}\) can be calculated as the minimal non-negative solution to
\begin{equation}
\label{eqn6}
(I_{u} - P_{U,U})\mathbf{t} = \mathbf{c}
\end{equation}
where \(\mathbf{c}\) is a column vector of length \(u\) whose entries are all \(1\)~\cite{grinstead-12}.

\subsection*{The Graph Labelling Semi-Supervised (GLaSS) Method}
\label{ss2.4}
Consider a graph \(G\), with \(u\) unlabelled nodes and \(\ell\) labelled nodes, and suppose that all labelled nodes have one of two labels; either \(K_{1}\) or \(K_{2}\). From the weighted adjacency matrix \(A\), construct the transition matrix \(P\), as in (\ref{eqn1}). Using \(P\), calculate the vector of expected times to absorption \(\mathbf{t}\), as in (\ref{eqn6}). The expected times to absorption may, optionally, be used as a filtering criterion; nodes with a large expected time to absorption, relative to the disibution of \(t_{i}\) over all nodes in the graph, may be excluded from further analysis. Once nodes have been optionally filtered using \(\mathbf{t}\), calculate the matrix of absorption probabilities \(H\), by (\ref{eqn4}), and calculate \(P(Y_{i} = K_{1})\) and \(P(Y_{i} = K_{2})\), for all \(i \in U\), as in (\ref{eqn5}). Because \(P(Y_{i} = K_{1}) + P(Y_{i} = K_{2}) = 1\), only one probability is required to classify the unlabelled nodes.

Once \(P(Y_{i}=K_{1})\) and \(P(Y_{i}=K_{2})\) have been obtained, these probabilities may be used to estimate labels for nodes in \(G\). For the purposes of this analysis, nodes are classified in the following way: Suppose that \(m\) of the \(u\) unlabelled nodes in \(G\) have a true label \(K_{2}\), and that the remaining \((u-m\)) unlabelled nodes have a true label \(K_{1}\). That is, the ratio of nodes with label \(K_{1}\) to nodes with label \(K_{2}\) is known, but which nodes should bear those labels is not. Sorting the probabilities \(P(Y_{i}=K_{1})\) from smallest to largest, the \(m\)\tss{th} order statistic (the \(m\)\tss{th} smallest probability) is chosen as a threshold \(\alpha\), and a binary classifier is implemented. If \(P(Y_{i} = K_{1}) > \alpha\), estimate the label for node \(v_{i}\) as \(K_{1}\); otherwise, if \(P(Y_{i} = K_{1}) \leq \alpha\), estimate the label for node \(v_{i}\) as \(K_{2}\). Thus, \(\alpha\) is chosen to assign a label of \(K_{1}\) to the \((u-m)\) nodes deemed most likely to have that label, and assigns a label of \(K_{2}\) to the remaining \(m\) unlabelled nodes in G.

Using this method, it is possible to estimate the label for every unlabelled node in \(G\). This method forms a modification and extension to the GLaSS method~\cite{glonek-18}, a graph labelling method in a semi-supervised setting. Hereafter, we refer to this modification as ``the GLaSS method''.

\section*{Data}
\label{s3}
Validating the GLaSS method requires graphs with a clear community structure and known labels for all nodes. To emulate a graph with few known labels, only a small subset of all known labels will be used by GLaSS, with the remaining labels withheld to emulate ``unlabelled'' nodes in the graph. All labels estimated by GLaSS can then be compared to actual, withheld labels, to assess performance. United States roll call voting data are chosen to validate the GLaSS method.

In the United States House of Representatives (the House) and the Senate, parliamentary procedure occasionally gives rise to roll call votes. In a roll call vote, the vote of every member of the House or the Senate is recorded, making it possible to see which members voted the same way. Roll call voting data for the House and the Senate can be modelled as an undirected graph, where each node represents a member of Congress, and a positive integer-weighted edge records the number of times respective members voted the same way. Roll call voting data for the House and the Senate are modelled as separate graphs.

The results of roll call votes in the House and the Senate for 42 separate Congresses, between 1935 and 2019,~\tss{a} have been collected for analysis, and modelled as 84 separate undirected graphs. The data has been made available on Voteview~\cite{vv-18} and Figshare for analysis.~\tss{b} For simplicity, in each Congress, the following rules are applied:
\begin{enumerate}
\item Only ``yea'' and ``nay'' votes are considered.
\item Only members whose party affiliation is Democrat or Republican are considered.
\item In cases where a member's party affiliation changes during a meeting of Congress, their party affiliation at the time they were elected is used.
\item In rare cases, a member of Congress does not sit for the entire meeting of Congress, and their seat is taken by a new member. In these cases, the voting records of both members are retained.~\tss{c}
\item In both the House and the Senate, votes where the Democrat and Republican leader cast the same vote (either ``yea'' or ``nay'') are not considered, as they provide no information about partisanship.
\item In both the House and the Senate, multiple members may serve (non-concurrently) as party leader. In these cases,  the vote cast by the party leader at the time the vote was held is considered when implementing rule 5.
\item In the House, multiple members may serve (non-concurrently) as Speaker of the House. In cases where only one Speaker is active for the meeting of Congress, votes cast by the Speaker are not considered.~\tss{d} In cases where multiple Speakers are active, votes cast by the first Speaker (chronologically) are not considered, but votes cast by subsequent Speakers are considered.~\tss{e} This rule does not apply to the President Pro Tempore of the Senate. 
\end{enumerate}

Because the party affiliation of each member of Congress is known, all nodes in each graph are labelled. For these analyses, only the labels of nodes corresponding to the Democrat and Republican Leaders are retained, thus all other nodes in each graph are ``unlabelled''. For example, the graph of the 74\tss{th} House comprises 429 nodes, but only three nodes (corresponding to the two Democrat leaders and one Republican leader) are labelled; the remaining 426 nodes in the graph are ``unlabelled''. Labels for these ``unlabelled'' nodes are used only to assess the accuracy of labels estimated by GLaSS. All graphs are either fully connected or nearly fully connected, and detailed summaries of each graph of the House and the Senate are contained in Tables~\ref{h-stats}~and~\ref{s-stats}, respectively.

\begin{table}[h!]
\caption{Years covered, total number of members (nodes), number of Democrats (excluding leaders), number of Republicans (excluding leaders), number of Democrat leaders, number of Republican leaders, and number of roll call votes for each House. Houses where the number of Democrats is shown in bold had a Democrat majority, and Houses where the number of Republicans is shown in bold had a Republican majority.}
\label{h-stats}
\centering
\begin{tabular}{cccccC{0.1mm}ccc}
\hline\noalign{\smallskip}
Congress & Years & \multicolumn{3}{c}{Members} & & \multicolumn{2}{c}{Leaders} & Votes \\
\cline{3-5}\cline{7-8}\noalign{\smallskip}
& & Total & Dem & Rep & & Dem & Rep & \\
\hline\noalign{\smallskip}
74 & 1935 - 37 & 429 & \textbf{322} & 104 & & 2 & 1 & 194 \\
75 & 1937 - 39 & 431 & \textbf{338} & 91 & & 1 & 1 & 135 \\
76 & 1939 - 41 & 452 & \textbf{272} & 177 & & 2 & 1 & 172 \\
77 & 1941 - 43 & 442 & \textbf{273} & 167 & & 1 & 1 & 95 \\
78 & 1943 - 45 & 445 & \textbf{226} & 217 & & 1 & 1 & 94 \\
79 & 1945 - 47 & 444 & \textbf{249} & 193 & & 1 & 1 & 149 \\
80 & 1947 - 49 & 446 & 193 & \textbf{251} & & 1 & 1 & 106 \\
81 & 1949 - 51 & 442 & \textbf{265} & 175 & & 1 & 1 & 169 \\
82 & 1951 - 53 & 445 & \textbf{238} & 205 & & 1 & 1 & 122 \\
83 & 1953 - 55 & 439 & 217 & \textbf{220} & & 1 & 1 & 63 \\
84 & 1955 - 57 & 437 & \textbf{233} & 202 & & 1 & 1 & 60 \\
85 & 1957 - 59 & 443 & \textbf{238} & 203 & & 1 & 1 & 106 \\
86 & 1959 - 61 & 441 & \textbf{284} & 155 & & 1 & 1 & 104 \\
87 & 1961 - 63 & 449 & \textbf{270} & 176 & & 2 & 1 & 123 \\
88 & 1963 - 65 & 442 & \textbf{260} & 180 & & 1 & 1 & 124 \\
89 & 1965 - 67 & 442 & \textbf{299} & 141 & & 1 & 1 & 251 \\
90 & 1967 - 69 & 437 & \textbf{247} & 188 & & 1 & 1 & 211 \\
91 & 1969 - 71 & 448 & \textbf{248} & 198 & & 1 & 1 & 139 \\
92 & 1971 - 73 & 443 & \textbf{257} & 184 & & 1 & 1 & 326 \\
93 & 1973 - 75 & 441 & \textbf{245} & 193 & & 1 & 2 & 562 \\
94 & 1975 - 77 & 441 & \textbf{293} & 146 & & 1 & 1 & 709 \\
95 & 1977 - 79 & 441 & \textbf{293} & 146 & & 1 & 1 & 783 \\
96 & 1979 - 81 & 440 & \textbf{279} & 159 & & 1 & 1 & 691 \\
97 & 1981 - 83 & 441 & \textbf{245} & 194 & & 1 & 1 & 354 \\
98 & 1983 - 85 & 439 & \textbf{271} & 166 & & 1 & 1 & 471 \\
99 & 1985 - 87 & 439 & \textbf{256} & 181 & & 1 & 1 & 554 \\
100 & 1987 - 89 & 440 & \textbf{260} & 178 & & 1 & 1 & 558 \\
101 & 1989 - 91 & 443 & \textbf{262} & 178 & & 2 & 1 & 521 \\
102 & 1991 - 93 & 440 & \textbf{269} & 169 & & 1 & 1 & 560 \\
103 & 1993 - 95 & 440 & \textbf{259} & 179 & & 1 & 1 & 735 \\
104 & 1995 - 97 & 444 & 207 & \textbf{235} & & 1 & 1 & 980 \\
105 & 1997 - 99 & 442 & 210 & \textbf{230} & & 1 & 1 & 794 \\
106 & 1999 - 01 & 438 & 212 & \textbf{224} & & 1 & 1 & 699 \\
107 & 2001 - 03 & 441 & 212 & \textbf{227} & & 1 & 1 & 530 \\
108 & 2003 - 05 & 439 & 207 & \textbf{230} & & 1 & 1 & 721 \\
109 & 2005 - 07 & 439 & 202 & \textbf{233} & & 1 & 3 & 714 \\
110 & 2007 - 09 & 448 & \textbf{241} & 205 & & 1 & 1 & 1233 \\
111 & 2009 - 11 & 446 & \textbf{262} & 182 & & 1 & 1 & 944 \\
112 & 2011 - 13 & 445 & 199 & \textbf{244} & & 1 & 1 & 1236 \\
113 & 2013 - 15 & 444 & 203 & \textbf{238} & & 1 & 2 & 888 \\
114 & 2015 - 17 & 440 & 189 & \textbf{249} & & 1 & 1 & 1028 \\
115 & 2017 - 19 & 450 & 199 & \textbf{249} & & 1 & 1 & 865 \\
\hline
\end{tabular}
\end{table}

\begin{table}[h!]
\caption{Years covered, total number of members (nodes), number of Democrats (excluding leaders), number of Republicans (excluding leaders), number of Democrat leaders, number of Republican leaders, and number of roll call votes for each Senate. Senates where the number of Democrats is shown in bold had a Democrat majority, and Senates where the number of Republicans is shown in bold had a Republican majority.}
\label{s-stats}
\centering
\begin{tabular}{cccccC{0.1mm}ccc}
\hline\noalign{\smallskip}
Congress & Years & \multicolumn{3}{c}{Members} & & \multicolumn{2}{c}{Leaders} & Votes \\
\cline{3-5}\cline{7-8}\noalign{\smallskip}
& & Total & Dem & Rep & & Dem & Rep & \\
\hline\noalign{\smallskip}
74 & 1935 - 37 & 97 & \textbf{71} & 24 & & 1 & 1 & 134 \\
75 & 1937 - 39 & 98 & \textbf{80} & 15 & & 2 & 1 & 130 \\
76 & 1939 - 41 & 100 & \textbf{72} & 26 & & 1 & 1 & 190 \\
77 & 1941 - 43 & 106 & \textbf{74} & 30 & & 1 & 1 & 122 \\
78 & 1943 - 45 & 102 & \textbf{59} & 40 & & 1 & 2 & 188 \\
79 & 1945 - 47 & 102 & \textbf{59} & 41 & & 1 & 1 & 151 \\
80 & 1947 - 49 & 97 & 45 & \textbf{50} & & 1 & 1 & 200 \\
81 & 1949 - 51 & 107 & \textbf{59} & 46 & & 1 & 1 & 375 \\
82 & 1951 - 53 & 98 & \textbf{50} & 45 & & 1 & 2 & 251 \\
83 & 1953 - 55 & 109 & 52 & \textbf{54} & & 1 & 2 & 159 \\
84 & 1955 - 57 & 98 & \textbf{50} & 46 & & 1 & 1 & 134 \\
85 & 1957 - 59 & 101 & \textbf{52} & 47 & & 1 & 1 & 149 \\
86 & 1959 - 61 & 103 & \textbf{66} & 35 & & 1 & 1 & 249 \\
87 & 1961 - 63 & 105 & \textbf{64} & 39 & & 1 & 1 & 272 \\
88 & 1963 - 65 & 102 & \textbf{68} & 32 & & 1 & 1 & 362 \\
89 & 1965 - 67 & 103 & \textbf{69} & 32 & & 1 & 1 & 295 \\
90 & 1967 - 69 & 101 & \textbf{63} & 36 & & 1 & 1 & 372 \\
91 & 1969 - 71 & 102 & \textbf{57} & 42 & & 1 & 2 & 328 \\
92 & 1971 - 73 & 99 & \textbf{54} & 43 & & 1 & 1 & 566 \\
93 & 1973 - 75 & 100 & \textbf{56} & 42 & & 1 & 1 & 582 \\
94 & 1975 - 77 & 99 & \textbf{60} & 37 & & 1 & 1 & 694 \\
95 & 1977 - 79 & 103 & \textbf{64} & 37 & & 1 & 1 & 495 \\
96 & 1979 - 81 & 100 & \textbf{58} & 40 & & 1 & 1 & 619 \\
97 & 1981 - 83 & 100 & 45 & \textbf{53} & & 1 & 1 & 471 \\
98 & 1983 - 85 & 101 & 45 & \textbf{54} & & 1 & 1 & 299 \\
99 & 1985 - 87 & 101 & 46 & \textbf{53} & & 1 & 1 & 355 \\
100 & 1987 - 89 & 101 & \textbf{54} & 45 & & 1 & 1 & 335 \\
101 & 1989 - 91 & 101 & \textbf{55} & 44 & & 1 & 1 & 280 \\
102 & 1991 - 93 & 102 & \textbf{57} & 43 & & 1 & 1 & 301 \\
103 & 1993 - 95 & 101 & \textbf{56} & 43 & & 1 & 1 & 465 \\
104 & 1995 - 97 & 103 & 47 & \textbf{53} & & 1 & 2 & 610 \\
105 & 1997 - 99 & 100 & 44 & \textbf{54} & & 1 & 1 & 329 \\
106 & 1999 - 01 & 102 & 45 & \textbf{55} & & 1 & 1 & 378 \\
107 & 2001 - 03 & 100 & 49 & \textbf{49} & & 1 & 1 & 332 \\
108 & 2003 - 05 & 99 & 47 & \textbf{50} & & 1 & 1 & 402 \\
109 & 2005 - 07 & 100 & 44 & \textbf{54} & & 1 & 1 & 376 \\
110 & 2007 - 09 & 101 & \textbf{49}\tss{*} & 50 & & 1 & 1 & 369 \\
111 & 2009 - 11 & 110 & \textbf{65} & 43 & & 1 & 1 & 512 \\
112 & 2011 - 13 & 101 & \textbf{52} & 47 & & 1 & 1 & 267 \\
113 & 2013 - 15 & 103 & \textbf{56} & 45 & & 1 & 1 & 433 \\
114 & 2015 - 17 & 98 & 43 & \textbf{53} & & 1 & 1 & 303 \\
115 & 2017 - 19 & 103 & 47 & \textbf{54} & & 1 & 1 & 378 \\
\hline
\end{tabular} \\
\tss{*} Democrats formed a working majority in coalition with two Independent Senators.
\end{table}

\section*{Results}
\label{s4}
Each House and Senate is modelled as a graph, and each graph is analysed using the GLaSS method, as described above. Expected time to absorption is calculated for each ``unlabelled'' node in each graph. Based on the distribution of \(t_{i}\), for each graph, no filtering is required, and labels are estimated for all ``unlabelled'' nodes in all graphs.

\begin{figure}[h!]
\caption{Clockwise from top right: histogram of absorption probabilities for the 110\tss{th} Senate; graph of the 110\tss{th} Senate; graph of the 90\tss{th} Senate; histogram of absorption probabilities for the 90\tss{th} Senate. In the histograms, red bars represent Republican members and blue bars represent Democrat members. In the graphs, red nodes represent Republican members and blue nodes represent Democrat members. Graphs are visualised using the Fruchterman-Reingold force-directed layout algorithm. The 90\tss{th} Senate was the most difficult to correctly label using GLaSS (F1 = 0.8571). This diffculty is reflected in the overlap of red and blue bars in the histogram (top left), and interspersal of red and blue nodes in the graph (bottom left). The isolated red node on the far right of the graph corresponds to the isolated red bar on the far right of the histogram. The 110\tss{th} Senate was perfectly labelled by GLaSS (F1 score = 1), as reflected by the clear separation of red and blue bars in the histogram (top right). Also of note is the clearer separation of red and blue nodes within the graph of the 110\tss{th} Senate (bottom right), suggesting a more obvious community structure.}
\includegraphics[width=120mm]{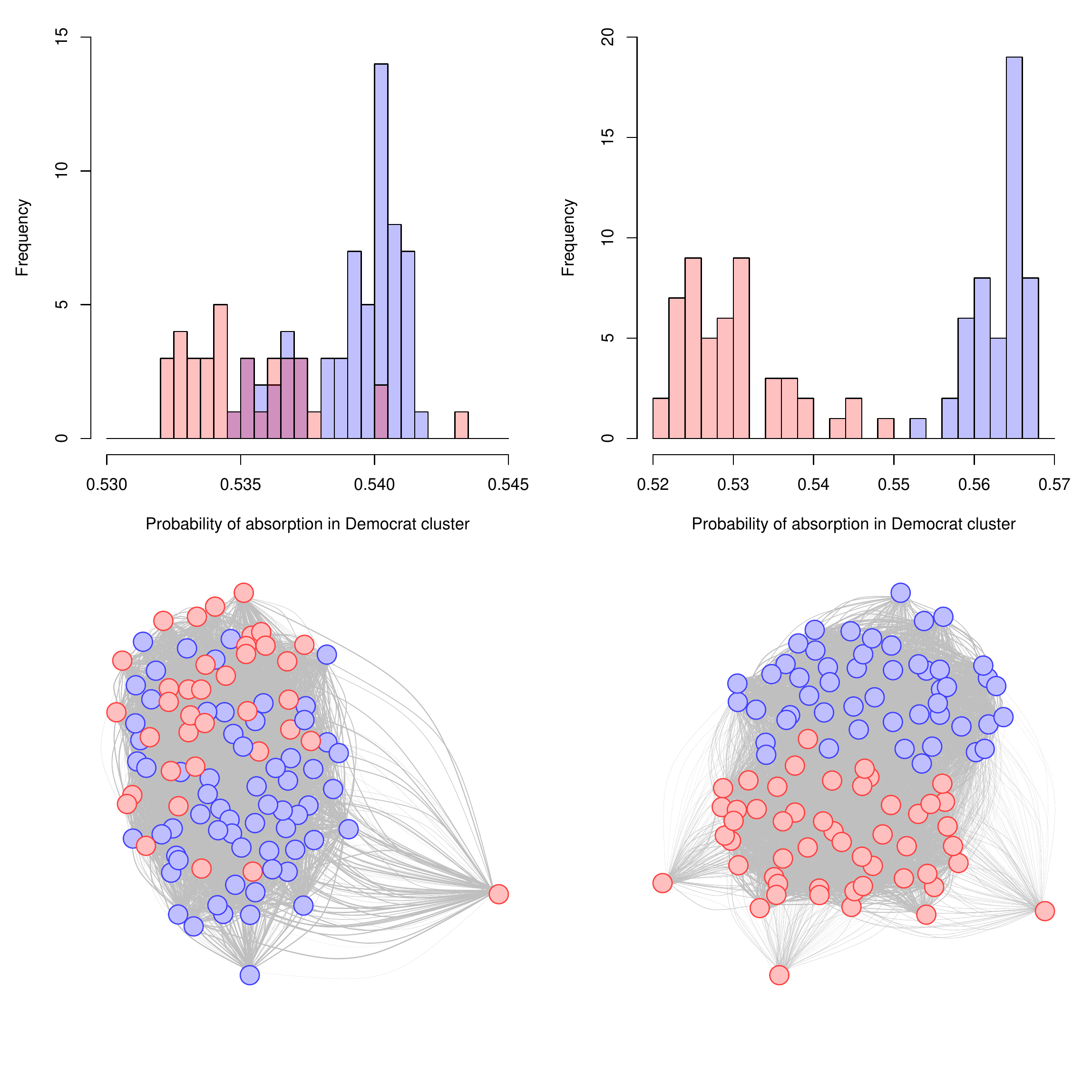}
\label{g-and-h}
\end{figure}

In graphs containing only two labelled nodes (one Democrat leader, one Republican leader), each labelled node forms an absorbing state, and the probability of being absorbed in the Democrat state of the corresponding DTMC is considered. In graphs containing more than two labelled nodes (multiple Democrat leaders or multiple Republican leaders), labelled nodes for each party are taken to form an absorbing class, and the probability of being absorbed in the Democrat class of the corresponding DTMC is considered. For illustrative purposes, full graphs, and histograms of absorption probabilities for the 90\tss{th} and 110\tss{th} Senates are provided in Figure~\ref{g-and-h}. Histograms for all Houses and all Senates show separation between Democrat and Republican members, though some overlap between clusters does exist for some Congresses.

\begin{figure}[h!]
\caption{Top: F1 Score for the House from 1935-37 (74\tss{th} House) to 2017-19 (115\tss{th} House). Bottom: Difference between the smallest standardised \(P(Y_{i}=Democrat)\) among all true Democrats and the largest standardised \(P(Y_{i}=Democrat)\) among all true Republicans for the House from 1935-37 (74\tss{th} House) to 2017-19 (115\tss{th} House). While there is some variability over time, F1 scores are relatively high for all Houses (minimum F1 score = 0.8776), indicating very strong performance of the GLaSS method in labelling members of the House as Democrat or Republican. In particular, every House from the 108\tss{th} (2003-05) onwards has an F1 score of 1, implying that the GLaSS method was able to perfectly identify the party affiliation of every member in those Houses. The plot of standardised differences shows the magnitude of overlap (values below the horizontal line at 0) or separation (values above the horizontal line at 0) between Democrats and Republicans, according to absorption probabilities calculated by the GLaSS method. F1 score appears to decrease with increasing magnitude of overlap, while also showing that the two parties have grown increasingly far apart since they first separated entirely in the 108\tss{th} House.}
\includegraphics[width=120mm]{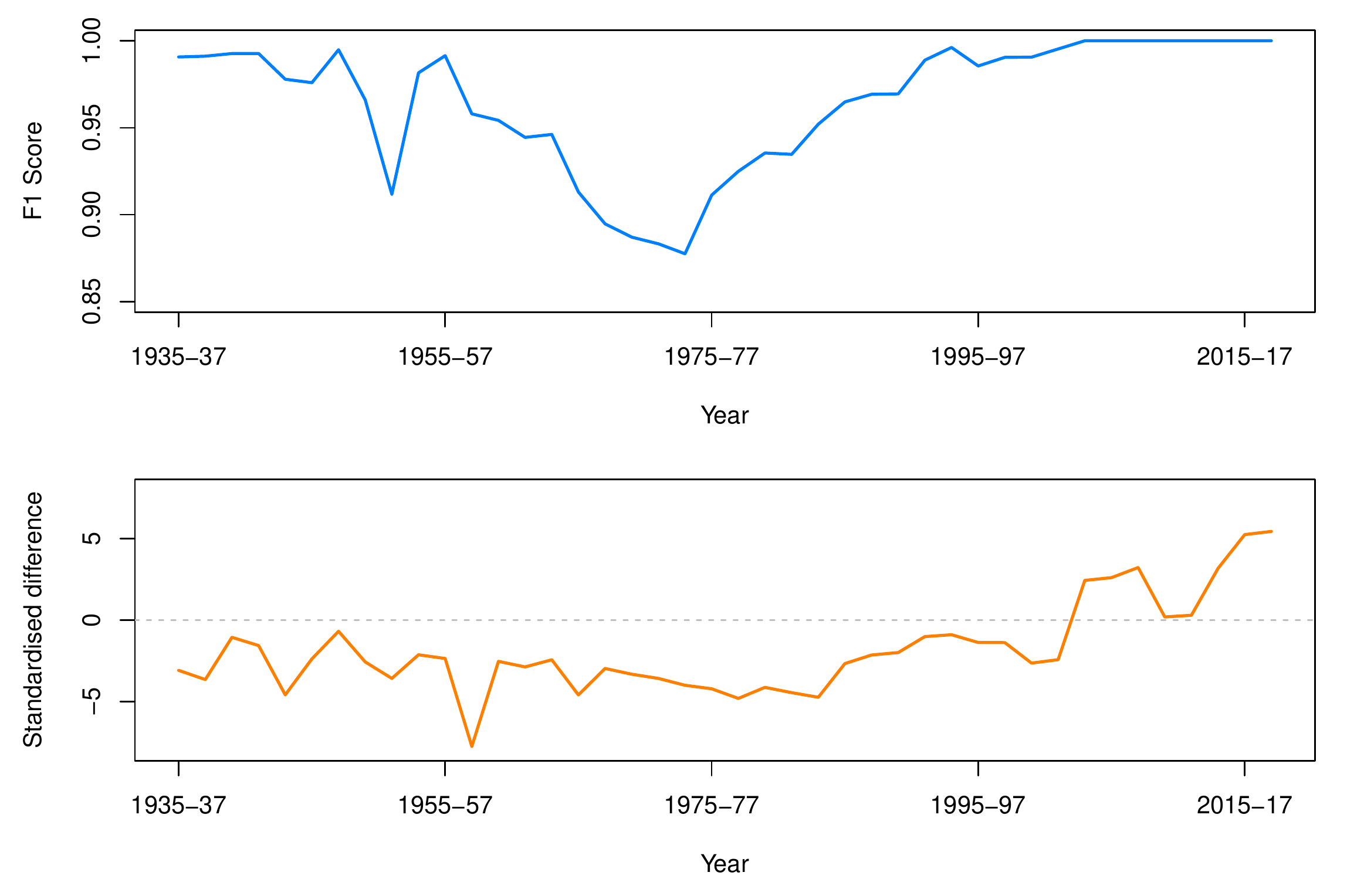}
\label{h-F1}
\end{figure}

\begin{figure}[h!]
\caption{Top: F1 Score for the Senate from 1935-37 (74\tss{th} Senate) to 2017-19 (115\tss{th} Senate). Bottom: Difference between the smallest standardised \(P(Y_{i}=Democrat)\) among all true Democrats and the largest standardised \(P(Y_{i}=Democrat)\) among all true Republicans for the Senate from 1935-37 (74\tss{th} Senate) to 2017-19 (115\tss{th} Senate). F1 scores show some variability over time, but are again relatively high for all Senates (minimum F1 score = 0.8571), indicating that the GLaSS method also performs very strongly in labelling members of the Senate as Democrat or Republican. Every Senate since the 110\tss{th} (2007-09) has an F1 score of 1, implying complete separation of the parties and perfect performance by GLaSS for those Senates. The Senate also experienced complete separation of Democrats and Republicans in 1983-85, 1995-97, and 1997-99 (98\tss{th}, 104\tss{th}, and 105\tss{th} Senates, respectively). The plot of standardised differences illustrates the magnitude of overlap or separation between Democrats and Republicans in the Senate, as measured by the GLaSS method. Some negative association between F1 score and magnitude of overlap is apparent, while it is also clear that the parties are now more separated in the Senate than at any time since 1935-37.}
\includegraphics[width=120mm]{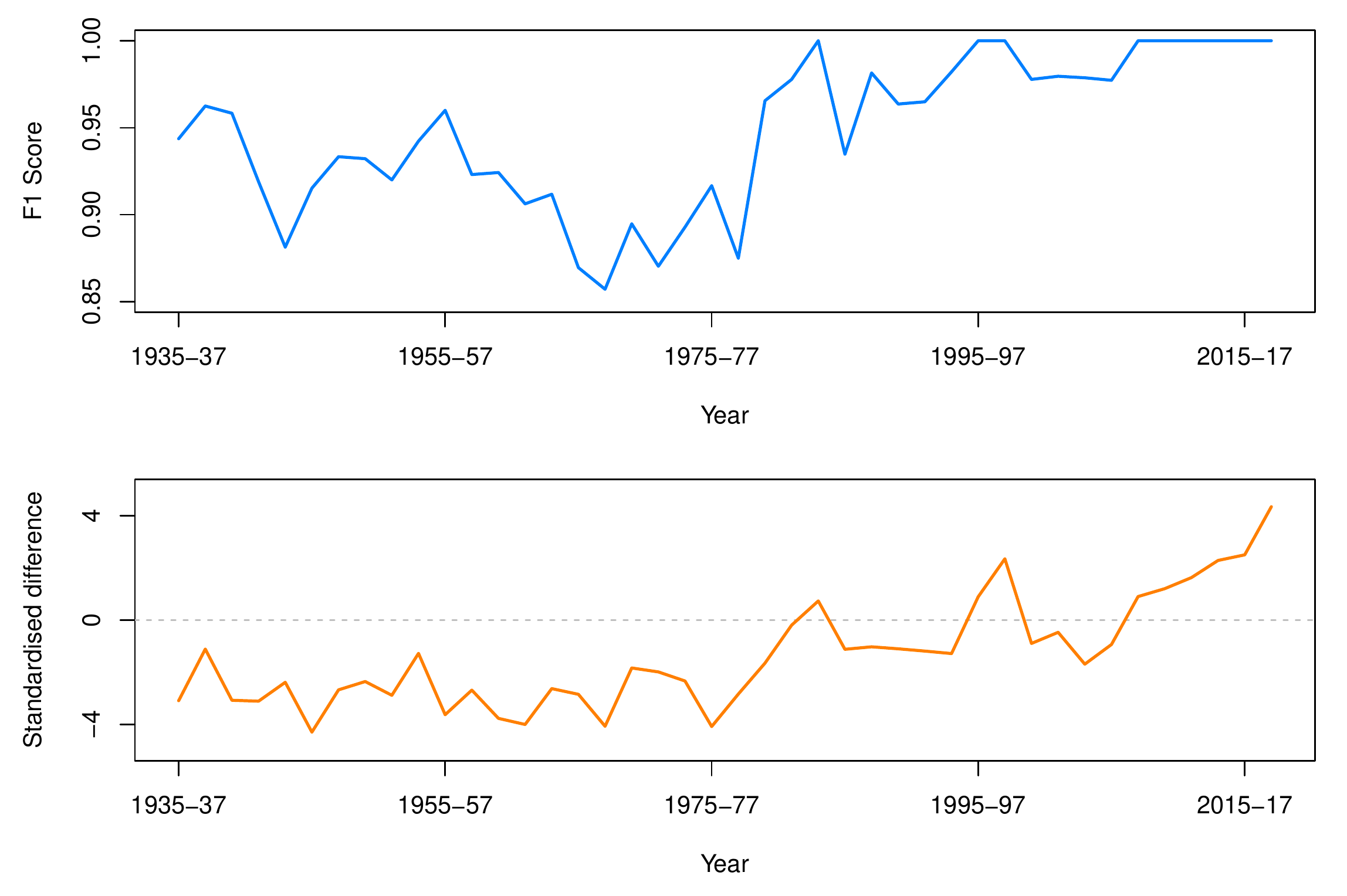}
\label{s-F1}
\end{figure}

Using the binary classifier in GLaSS, a threshold \(\alpha\) is chosen for each House and each Senate. If \(P(Y_{i} = Democrat) > \alpha\), then member \(i\) is labelled a Democrat; otherwise, member \(i\) is labelled a Republican. Estimated labels are compared to the true party affiliation for all ``unlabelled'' nodes in each graph. A confusion matrix is constructed for each graph, and used to calculate an F1 score, to measure the performance of GLaSS. An F1 score of 1 implies that GLaSS is able to correctly label all ``unlabelled'' members in the corresponding House or Senate. Plots of F1 score over time are given for the House and the Senate in Figures~\ref{h-F1}~and~\ref{s-F1}, respectively. F1 scores calculated for the 84 graphs range from a minimum of 0.8571 (achieved by the 90\tss{th} Senate) to a maximum of 1 (achieved by 8 Houses and 9 Senates).

To better understand the behaviour of graphs with an F1 score of 1, absorption probabilities are standardised using the population mean and pooled variance. The difference between the lowest standardised \(P(Y_{i}=Democrat)\) among all true Democrats and the highest standardised \(P(Y_{i}=Democrat)\) among all true Republicans is calculated and plotted against time (see Figures~\ref{h-F1}~and~\ref{s-F1}). In conjunction with plots of F1 score over time, these figures illustrate where overlap exists between Democrats and Republicans, but also by how much Democrats and Republicans are separated when they do not overlap.

Figures~\ref{h-F1} and \ref{s-F1} show a notable drop in F1 score through the 1960s and 1970s, corresponding to a period where the GLaSS method is less able to determine the party affiliation of members of the House and the Senate. The causes for this drop are unclear, but represent a possible decrease in partisanship during this period. The figures also show that, over the last 10 to 15 years, partisanship in both the House and the Senate has increased significantly. During this period, the F1 scores for both the House and the Senate are constant at 1, and plots of standardised difference over time show increasing separation between Democrats and Republicans, as measured by the GLaSS method.

\subsection*{Sensitivity Analysis}
\label{ss4.1}
The application of data cleaning rules, described in Data, requires special care, to ensure results are not unduly affected. While some rules are applied only to clarify the structure of roll call voting networks, rules 1, 2, and 5 remove observations from the datasets used to construct each graph. To determine what impact, if any, the application of rules 1, 2, and 5 have on results, sensitivity and exploratory data analyses are performed. All data cleaning is performed sequentially; that is, rule 1 is applied before rule 2, rule 2 is applied before rule 3, etc.

Before the application of any data cleaning rules, each dataset (raw data) contains all members in the House or Senate who cast at least one vote (of any kind, not strictly ``yea'' or ``nay'') during the meeting of that Congress, and every vote cast by every member is recorded. Rule 1 removes all recorded votes that are not strictly ``yea'' or ``nay''. This is done to avoid misinterpreting the intention behind a vote that is not actively in favour of (``yea'') or actively against (``nay'') a proposed bill. For example, in the event that a member expects to be absent during a vote, they may informally ``pair'' with a member from the opposite side who will be present, forming an agreement under which neither the absent member nor the present member register a formal vote of ``yea'' or ``nay'', though the ``pair'' is still recorded. Rule 2 removes all members whose party is not Democrat or Republican. This is done to avoid attempting to label Independent members and members of minor parties, who do not have leaders (or, in the case of Independents, party colleagues), and who cannot be correctly labelled as either Democrat or Republican.

Applying rule 1 to the raw data for the House and Senate does remove some members and some recorded votes. On average, the application of rule 1 reduces the number of members in the House by 0.1436\%, and the number of recorded votes by 11.2780\%. In the most extreme cases, the number of members in the 75\tss{th} house is reduced by 0.6696\% (448 members reduced to 445), and the number of recorded votes in the 79\tss{th} House is reduced by 23.9810\% (100,271 votes reduced to 76,225). On average, the application of rule 1 reduces the number of members in the Senate by 0.2448\%, and the number of recorded votes by 12.3591\%. In the most extreme cases, the number of members in the 79\tss{th} Senate is reduced by 5.5046\% (109 members reduced to 103), and the number of recorded votes in the 81\tss{st} Senate is reduced by 29.5413\% (49,595 votes reduced to 34,944). While individual members and the votes of individual members are removed by applying rule 1, no roll call votes are excluded.

Rule 2 removes further members and recorded votes. On average, the application of rule 2 further reduces the number of members in the House by 0.5348\%, and the number of recorded votes by 0.3800\%. In the most extreme cases, the number of members in the 75\tss{th} House is further reduced by 3.1461\% (445 members reduced to 431), and the number of recorded votes in the 75\tss{th} House is further reduced by 3.3199\% (54,911 votes reduced to 53,088). On average, the application of rule 2 further reduces the number of members in the Senate by 1.6379\%, and the number of recorded votes by 1.1547\%. In the most extreme cases, the number of members in the 75\tss{th} Senate is further reduced by 3.9216\% (102 members reduced to 98), and the number of recorded votes in the 76\tss{th} Senate is further reduced by 4.1056\% (19,729 votes reduced to 18,919). As with rule 1, rule 2 removes indivudual members and the votes of individual members, but does not remove any roll call votes.

From these exploratory data analyses, it is clear that rules 1 and 2 do not significantly reduce the number of nodes (members) in any roll call voting network. While the application of rule 1 does significantly reduce the total number of recorded votes in some Houses and Senates, the application of this rule is justified by the removal of ambiguous votes (neither ``yea'' nor ``nay'').

This study examines partisanship trends within the House and Senate. However, applying rule 5 to roll call voting data explicitly removes votes where the Democrat leader and Republican leader vote the same way (either both ``yea'', or both ``nay). That is, votes that attract a bipartisan response (support or rejection by both parties) are not included in the roll call voting networks analysed, under the assumption that they do not contribute meaningful information about the party affiliation of members. To ensure that the exclusion of these votes does not significantly affect the results of analyses using the GLaSS method, two roll call voting networks are re-analysed with agreeing votes included (votes where the Democrat and Republican leaders vote the same way). The least and most partisan roll call voting networks are chosen for analysis. By F1 score, the 90\tss{th} Senate is the least partisan network, and, by standardised difference, the 115\tss{th} House is the most partisan network.

Analysing the original 90\tss{th} Senate and 115\tss{th} House networks yields respective F1 scores of 0.8517 and 1. With agreeing votes included in the roll call voting networks, analsyis with GLaSS again yields F1 scores of 0.8571 and 1, respectively. This indicates that GLaSS performs equally well with or without the inclusion of agreeing votes in roll call voting networks, and that agreeing votes contribute no meaningful information overall about the party affiliation of members, or partisanship within the House and Senate.

\begin{figure}[h!]
\caption{Clockwise from top right: histogram of absorption probabilities for the 115\tss{th} House, without agreeing votes; histogram of absorption probabilities for the 115\tss{th} House, with agreeing votes; scatterplot of absorption probabilities for the 115\tss{th} House; scatterplot of absorption probabilities for the 90\tss{th} Senate; histogram of absorption probabilities for the 90\tss{th} Senate, with agreeing votes; histogram of absorption probabilities for the 90\tss{th} Senate, without agreeing votes. In the histograms, red bars represent Republicans and blue bars represent Democrats. In the scatterplots, red crosses represent Republicans and blue crosses represent Democrats. The 90\tss{th} Senate and 115\tss{th} House represent the least partisan and most partisan roll call voting networks analysed, respectively. For a detailed interpretation of these plots, see Sensitivity Analysis.}
\includegraphics[width=120mm]{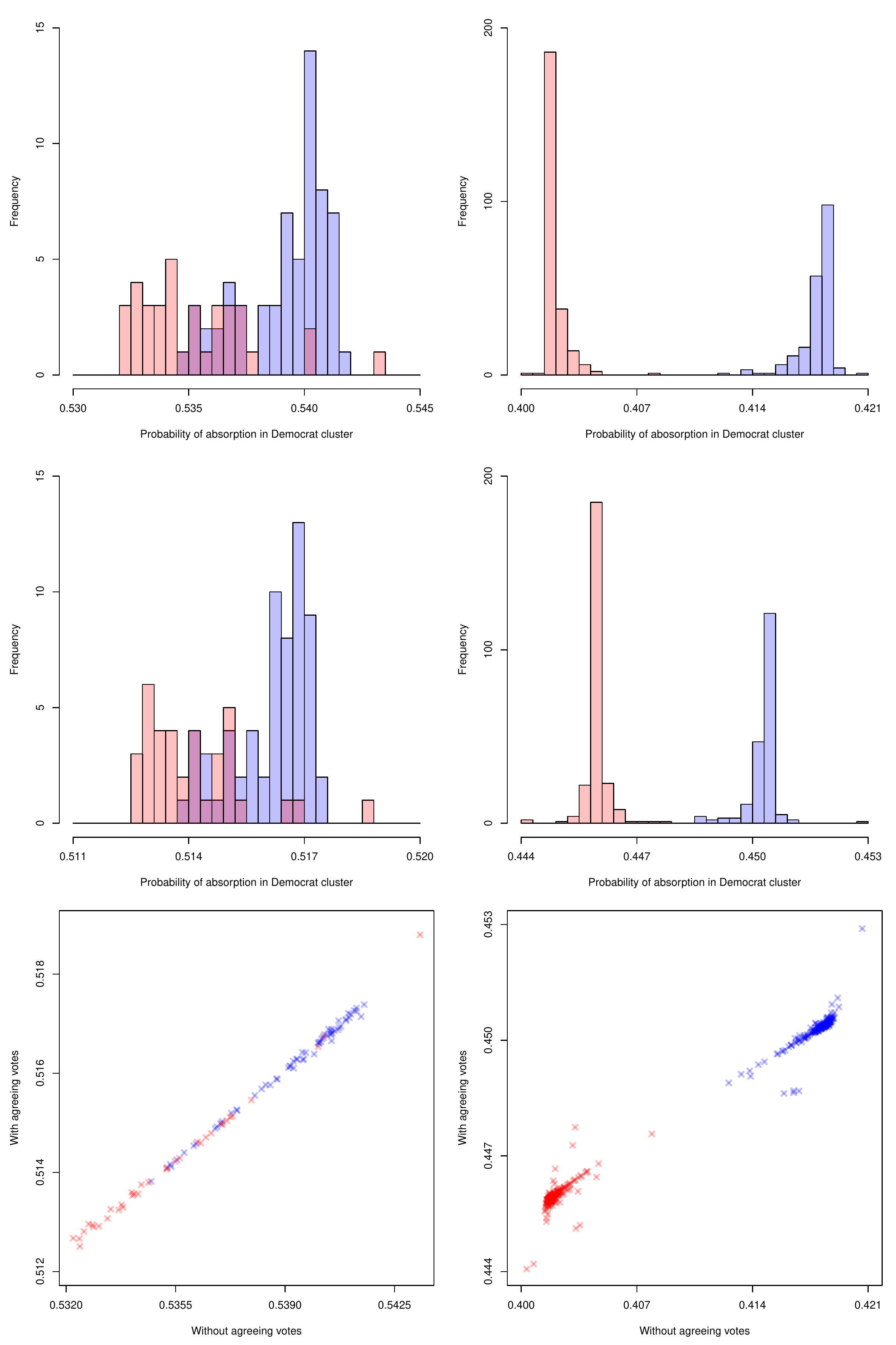}
\label{sens}
\end{figure}

Figure~\ref{sens} provides a visualisation of the effect of including agreeing votes on the performance of the GLaSS method. The top two histograms show the original calculated absorption probabilities for the 90\tss{th} Senate and 115\tss{th} House (without agreeing votes), while histograms in the middle row show the calculated absorption probabilities for both Congress with agreeing votes included. From the histograms, it is clear that the inclusion of agreeing votes has affected the distribution of absorption probabilities for both Congreses. Specifically, including agreeing votes has shifted the location of both distributions towards 0.5, and reduced the spread of each distribution. However, considering pairs of histograms shows that changes in the distribution of absorption probabilities for each Congress has not significantly affected the level of overlap or separation between Democrats and Republicans. In particular, histograms for the 90\tss{th} Senate (top left and middle left) show that the level of overlap between parties has not significantly changed, while histograms for the 115\tss{th} House (top right and middle right) both show that Democrats and Republicans are entirely separated.

Scatterplots at the bottom of Figure~\ref{sens} show calculated absorption probabilities for each member, with (y-axis) and without (x-axis) agreeing votes,  in the 90\tss{th} Senate (bottom left) and 115\tss{th} House (bottom right), respectively. The scatterplot for the 90\tss{th} Senate shows very strong linearity, indicating an almost perfect correspondence between the ordering of members (by calculated absorption probability) with and without agreeing votes. Pearson and Spearman correlations for the 90\tss{th} Senate are 0.9986 and 0.9978, respectively, supporting the notion that including agreeing votes has minimal impact on analyses with the GLaSS method. The scatterplot for the 115\tss{th} House shows weaker linearity, but Pearson and Spearman correlations of 0.9946 and 0.9581, respectively, again support the notion that agreeing votes do not significantly affect results. Further, as in the histograms for the 115\tss{th} House, Democrats and Republicans are completely separated, both vertically and horizontally, in the scatterplot. Based on the results of these sensitivity and exploratory data analyses, there is no evidence that the data cleaning rules applied to US roll call voting networks have significantly impacted analyses with the GLaSS method, or biased results towards detecting partisanship trends within the House and Senate.

\subsection*{The Effects of Party Affiliation and Control on Partisanship}
\label{ss4.2}
To better understand the factors that may influence partisanship within the House and the Senate, as measured by F1 scores calculated for the GLaSS method, two regression models are fitted. In particular, the models are fitted to investigate whether partisanship in the House and Senate (as measured by F1 score) can be reliably predicted by simple, extraneous factors, or whether the findings of analysis using the GLaSS method are non-trivial. The first model examines the effect of three factors on F1 score; which party holds a majority in the House, which party holds a majority in the Senate, and which party holds the Presidency. Each factor comprises two levels - in each instance, Democrats or Republicans are in majority in the House, Democrats or Republicans are in majority in the Senate, and the sitting President is a Democrat or Republican - and all two-way and three-way interaction terms are considered. Thus, the full model is
\begin{equation}
\label{model1}
F \sim H + S + P + (H \times S) + (H \times P) + (S \times P) + (H \times S \times P),
\end{equation}
where \(F\) is F1 score, \(H\) is the party in majority in the House, \(S\) is the party in majority in the Senate, \(P\) is the party affiliation of the sitting President, and \(\times\) denotes an interaction between the categorical variables. The model is fitted for F1 scores from the House and the Senate separately, but no significant predictors are identified in either case.

In the second model, the number of factors is reduced. For the House, two binary factors are considered; whether the party that holds a majority in the Senate is the same as the party that holds a majority in the House, and whether the party that holds the Presidency is the same as the party that holds a majority in the House. An equivalent model is specified for the Senate, and two-way interactions are considered in both cases. For the House, the full model is
\begin{equation}
\label{h-model2}
F \sim S' + P' + (S' \times P'),
\end{equation}
where \(S'\) and \(P'\) denote whether the party in control of the Senate and the Presidency, respectively, is the same as the party in control of the House. For the Senate, the full model is
\begin{equation}
\label{s-model2}
F \sim H' + P' + (H' \times P'),
\end{equation}
where \(H'\) and \(P'\) denote whether the party in control of the House and the Presidency, respectively, is the same as the party in control of the Senate. In (\ref{h-model2}) and (\ref{s-model2}), \(F\) denotes F1 score for the House and Senate, respectively, and \(\times\) is as previously defined. Again, no significant predictors of F1 are identified for the House or the Senate. Thus, while partisanship has clearly varied over time in both the House and the Senate, it appears that this variation is not explained by which party controls each branch of the US Government, or the interplay between controlling parties, lending credence to the notion that the findings about partisanship, by analysis using the GLaSS method, are non-trivial.

\section*{Conclusions}
\label{s5}
Graph labelling is a fundamental task within network science, with diverse applications. This work builds upon a previously introduced~\cite{glonek-18} semi-supervised graph labelling method using random walks to absorption, the GLaSS method, and uses it to analyse a collection of undirected political networks from the United States House of Representatives and Senate. In these networks, the GLaSS method is used to estimate the party affiliation of members of the House and the Senate based on roll call voting data. The GLaSS method shows universally strong performance in analysing these networks, returning F1 scores in excess of 0.85 for all 84 graphs, even where graphs display significant overlap between the two communities.

In 20\% of cases (17 of the 84 networks analysed), GLaSS returns an F1 score of 1, indicating perfect labelling of all members in the House or the Senate based on random walks to absorption in these roll call voting networks. Previous work evaluating GLaSS showed that it outperormed other supervised and unsupervised random walk-based graph labelling methods in analysing a small series of undirected political networks~\cite{glonek-18}. Results in this work provide more evidence that continued investigation and evaluation of the GLaSS method is warranted. Future work will extend this work to examine the performance of the GLaSS method for graphs of varying size, connectedness, density, and with different numbers of known labels. Extending the GLaSS method to label graphs in the general \(k\)-community setting, and graphs with fewer labelled nodes than clusters, is of particular interest, as is benchmarking the performance of the GLaSS method against other graph labelling methods.

This work also provides reinforcing empirical evidence of increasing partisanship in United States politics~\cite{andris-15,poole-84}, while also representing the first such analysis to use a random walk-based graph labelling method. Analysis of roll call voting data, in both the House and the Senate, using GLaSS, shows that the Democrat and Republican parties have undergone a recent rapid separation. The party affiliation of members can now be entirely predicted by their voting trends, where previously some uncertainty existed. This raises important questions about the causes of the recent and increasing polarisation in both the House and the Senate, as well as what factors historically decreased political partisanship in Congresses with lower F1 scores as calculated by GLaSS. Regression modelling indicates that variations in partisanship cannot be explained by which parties control the House, the Senate, and the Presidency.

In an applied setting, future work will also use GLaSS to further explore social, political, and other networks. Online and social-media networks are of particular interest, with a growing body of work examining the structure, dynamics, and polarisation of online social networks~\cite{fish-16,garimella-17,rizoiu-18,shai-17}. Future applied work with GLaSS will examine these characteristics for new and existing graphs. The roll call voting data presented here have a clear longitudinal structure; the construction of a metagraph from graphs of individual Houses or Senates and extension of the GLaSS method to analyse metagraphs is also an area for future work.

\section*{Endnotes}
\hspace*{1em}\tss{a} Each meeting of Congress begins on January 3 and runs for a period of two years. \\
\hspace*{1em}\tss{b} Primary data is available from Voteview \(<\)\emph{https://voteview.com/data}\(>\), with supporting datasets and documented code made available on Figshare \linebreak \(<\)\emph{https://adelaide.figshare.com/articles/ANS\_CN2018\_Special\_Issue/7869311}\(>\). \\
\hspace*{1em}\tss{c} Consequently, while the House has 435 seats and the Senate has 100 seats, graphs may contain more than 435 and 100 nodes, respectively. \\
\hspace*{1em}\tss{d} Conventionally, the Speaker of the House participates in very few votes. \\
\hspace*{1em}\tss{e} Prior to serving as Speaker, such members are generally party leaders, and at least active voting members of Congress.


\section*{Declaration}
\begin{backmatter}
\section*{Acknowledgements}
MG acknolwedges the support received through the provision of scholarships by the University of Adelaide and Data to Decisions CRC. All authors thank Data to Decisions CRC and the ARC Centre of Excellence for Mathematical and Statistical Frontiers for their financial support.

\section*{Authors' contributions}
All authors participated in the conception and design of this work. MG prepared and analyzed the data and drafted the manuscript. All authors critically reviewed the manuscript and approved its final form. All authors read and approved the final manuscript.

\section*{Availability of data and materials}
The code and datasets generated and analysed during the current study are available on Voteview~\cite{vv-18} and Figshare.~\tss{b}

\section*{Competing interests}
The authors declare that they have no competing interests.


\bibliographystyle{bmc-mathphys} 

\end{backmatter}
\end{document}